

\def\singlespace{\normalbaselines}
\def\oneandahalfspace{\baselineskip=1.15\normalbaselineskip plus 1pt
\lineskip=2pt\lineskiplimit=1pt}

\def\np{\vfill\eject}
\def\nl{\hfil\break}

\def\nofirstpagenoten{\nopagenumbers\footline={\ifnum\pageno>1\tenrm
\hss\folio\hss\fi}}
\def\nofirstpagenotwelve{\nopagenumbers\footline={\ifnum\pageno>1\twelverm
\hss\folio\hss\fi}}
\def\leaderfill{\leaders\hbox to 1em{\hss.\hss}\hfill}
\def\ft#1#2{{\textstyle{{#1}\over{#2}}}}
\def\frac#1/#2{\leavevmode\kern.1em
\raise.5ex\hbox{\the\scriptfont0 #1}\kern-.1em/\kern-.15em
\lower.25ex\hbox{\the\scriptfont0 #2}}
\def\sfrac#1/#2{\leavevmode\kern.1em
\raise.5ex\hbox{\the\scriptscriptfont0 #1}\kern-.1em/\kern-.15em
\lower.25ex\hbox{\the\scriptscriptfont0 #2}}


\parindent=20pt
\def\narrow{\advance\leftskip by 40pt \advance\rightskip by 40pt}

\def\AB{\bigskip
        \centerline{\bf ABSTRACT}\medskip\narrow}
\def\nonarrower{\advance\leftskip by -40pt\advance\rightskip by -40pt}
\def\AE{\bigskip\nonarrower}

\def\boxit#1{\vbox{\hrule\hbox{\vrule\kern3pt
        \vbox{\kern3pt#1\kern3pt}\kern3pt\vrule}\hrule}}

\def\gtorder{\mathrel{\raise.3ex\hbox{$>$}\mkern-14mu
             \lower0.6ex\hbox{$\sim$}}}
\def\ltorder{\mathrel{\raise.3ex\hbox{$<$}|mkern-14mu
             \lower0.6ex\hbox{\sim$}}}
\def\dalemb#1#2{{\vbox{\hrule height .#2pt
        \hbox{\vrule width.#2pt height#1pt \kern#1pt
                \vrule width.#2pt}
        \hrule height.#2pt}}}

\font\fourteentt=cmtt10 scaled \magstep2
\font\fourteenbf=cmbx12 scaled \magstep1
\font\fourteenrm=cmr12 scaled \magstep1
\font\fourteeni=cmmi12 scaled \magstep1
\font\fourteenss=cmss12 scaled \magstep1
\font\fourteensy=cmsy10 scaled \magstep2
\font\fourteensl=cmsl12 scaled \magstep1
\font\fourteenex=cmex10 scaled \magstep2
\font\fourteenit=cmti12 scaled \magstep1
\font\twelvett=cmtt10 scaled \magstep1 \font\twelvebf=cmbx12
\font\twelverm=cmr12 \font\twelvei=cmmi12
\font\twelvess=cmss12 \font\twelvesy=cmsy10 scaled \magstep1
\font\twelvesl=cmsl12 \font\twelveex=cmex10 scaled \magstep1
\font\twelveit=cmti12
\font\tenss=cmss10
 
 \font\ninebf=cmbx7 scaled \magstep1
\font\ninerm=cmr7 scaled \magstep1 \font\ninei=cmmi7 scaled \magstep1
\font\ninesy=cmsy7 scaled \magstep1 
\font\eightrm=cmr7 scaled 1140 
 
\font\sevenbf=cmbx7 \font\sevenrm=cmr7 \font\seveni=cmmi7
\font\sevensy=cmsy7 

\catcode`@=11
\newskip\ttglue
\newfam\ssfam

\def\fourteenpoint{\def\rm{\fam0\fourteenrm}
\textfont0=\fourteenrm \scriptfont0=\tenrm \scriptscriptfont0=\sevenrm
\textfont1=\fourteeni \scriptfont1=\teni \scriptscriptfont1=\seveni
\textfont2=\fourteensy \scriptfont2=\tensy \scriptscriptfont2=\sevensy
\textfont3=\fourteenex \scriptfont3=\fourteenex \scriptscriptfont3=\fourteenex
\def\it{\fam\itfam\fourteenit} \textfont\itfam=\fourteenit
\def\sl{\fam\slfam\fourteensl} \textfont\slfam=\fourteensl
\def\bf{\fam\bffam\fourteenbf} \textfont\bffam=\fourteenbf
\scriptfont\bffam=\tenbf \scriptscriptfont\bffam=\sevenbf
\def\tt{\fam\ttfam\fourteentt} \textfont\ttfam=\fourteentt
\def\ss{\fam\ssfam\fourteenss} \textfont\ssfam=\fourteenss
\tt \ttglue=.5em plus .25em minus .15em
\normalbaselineskip=16pt
\abovedisplayskip=16pt plus 4pt minus 12pt
\belowdisplayskip=16pt plus 4pt minus 12pt
\abovedisplayshortskip=0pt plus 4pt
\belowdisplayshortskip=9pt plus 4pt minus 6pt
\parskip=5pt plus 1.5pt
\setbox\strutbox=\hbox{\vrule height12pt depth5pt width0pt}
\let\sc=\tenrm
\let\big=\fourteenbig \normalbaselines\rm}
\def\fourteenbig#1{{\hbox{$\left#1\vbox to12pt{}\right.\n@space$}}}

\def\twelvepoint{\def\rm{\fam0\twelverm}
\textfont0=\twelverm \scriptfont0=\ninerm \scriptscriptfont0=\sevenrm
\textfont1=\twelvei \scriptfont1=\ninei \scriptscriptfont1=\seveni
\textfont2=\twelvesy \scriptfont2=\ninesy \scriptscriptfont2=\sevensy
\textfont3=\twelveex \scriptfont3=\twelveex \scriptscriptfont3=\twelveex
\def\it{\fam\itfam\twelveit} \textfont\itfam=\twelveit
\def\sl{\fam\slfam\twelvesl} \textfont\slfam=\twelvesl
\def\bf{\fam\bffam\twelvebf} \textfont\bffam=\twelvebf
\scriptfont\bffam=\ninebf \scriptscriptfont\bffam=\sevenbf
\def\tt{\fam\ttfam\twelvett} \textfont\ttfam=\twelvett
\def\ss{\fam\ssfam\twelvess} \textfont\ssfam=\twelvess
\tt \ttglue=.5em plus .25em minus .15em
\normalbaselineskip=14pt
\abovedisplayskip=14pt plus 3pt minus 10pt
\belowdisplayskip=14pt plus 3pt minus 10pt
\abovedisplayshortskip=0pt plus 3pt
\belowdisplayshortskip=8pt plus 3pt minus 5pt
\parskip=3pt plus 1.5pt
\setbox\strutbox=\hbox{\vrule height10pt depth4pt width0pt}
\let\sc=\ninerm
\let\big=\twelvebig \normalbaselines\rm}
\def\twelvebig#1{{\hbox{$\left#1\vbox to10pt{}\right.\n@space$}}}

\def\tenpoint{\def\rm{\fam0\tenrm}
\textfont0=\tenrm \scriptfont0=\sevenrm \scriptscriptfont0=\fiverm
\textfont1=\teni \scriptfont1=\seveni \scriptscriptfont1=\fivei
\textfont2=\tensy \scriptfont2=\sevensy \scriptscriptfont2=\fivesy
\textfont3=\tenex \scriptfont3=\tenex \scriptscriptfont3=\tenex
\def\it{\fam\itfam\tenit} \textfont\itfam=\tenit
\def\sl{\fam\slfam\tensl} \textfont\slfam=\tensl
\def\bf{\fam\bffam\tenbf} \textfont\bffam=\tenbf
\scriptfont\bffam=\sevenbf \scriptscriptfont\bffam=\fivebf
\def\tt{\fam\ttfam\tentt} \textfont\ttfam=\tentt
\def\ss{\fam\ssfam\tenss} \textfont\ssfam=\tenss
\tt \ttglue=.5em plus .25em minus .15em
\normalbaselineskip=12pt
\abovedisplayskip=12pt plus 3pt minus 9pt
\belowdisplayskip=12pt plus 3pt minus 9pt
\abovedisplayshortskip=0pt plus 3pt
\belowdisplayshortskip=7pt plus 3pt minus 4pt
\parskip=0.0pt plus 1.0pt
\setbox\strutbox=\hbox{\vrule height8.5pt depth3.5pt width0pt}
\let\sc=\eightrm
\let\big=\tenbig \normalbaselines\rm}
\def\tenbig#1{{\hbox{$\left#1\vbox to8.5pt{}\right.\n@space$}}}
\let\rawfootnote=\footnote \def\footnote#1#2{{\rm\parskip=0pt\rawfootnote{#1}
{#2\hfill\vrule height 0pt depth 6pt width 0pt}}}

\def\tenfoot{\tenpoint\hskip-\parindent\hskip-.1cm}

\overfullrule=0pt
\twelvepoint
\def\sbullet{\raise.2em\hbox{$\scriptscriptstyle\bullet$}}
\nofirstpagenotwelve
\hsize=16.5 truecm
\baselineskip 15pt

\def\ft#1#2{{\textstyle{{#1}\over{#2}}}}

\def\a{\alpha_0}

\def\del{\partial}

\def\oneone{\rlap 1\mkern4mu{\rm l}}

\oneandahalfspace
\rightline{CTP TAMU-22/92}
\rightline{April 1992}

\vskip 2truecm
\centerline{\bf On Realisations of $W$ Algebras}
\vskip 1.5truecm
\centerline{H. Lu and C.N. Pope\footnote{}{\tenfoot Supported in part
by the U.S. Department of Energy, under
grant DE-FG05-91ER40633.}}
\vskip 1.5truecm
\centerline{\it Center
for Theoretical Physics,
Texas A\&M University,}
\centerline{\it College Station, TX 77843--4242, USA.}

\vskip 1.5truecm
\AB\singlespace

     It has been known for some time that $W$ algebras can be realised in
terms of an energy-momentum tensor together with additional free scalar
fields.  Some recent results have shown that more general realisations are
also possible.  In this paper, we consider a wide class of realisations that
may be obtained from the Miura transformation, related to the existence of
canonical subalgebras of the Lie algebras on which the $W$ algebras are
based.  We give explicit formulae for all realisations of this kind, and
discuss their applications in $W$-string theory.

\AE\oneandahalfspace

\vskip 2truecm
\centerline{\tenpoint Available from hep-th/9204038}

\np

\noindent{\it 1) Introduction}
\bigskip

      Two-dimensional conformal field theories with non-linear local
symmetry algebras, name- ly $W$ algebras, have applications in extensions of
string theories [1-6].  The physical interpretation of the corresponding
string theory depends upon the specific realisation of the $W$ symmetry.
Since the $W$ algebras, which are higher-spin extensions of the Virasoro
algebra, are generically non-linear, realisations are not easy to come by.
Unlike the Virasoro algebra, one cannot obtain new realisations simply by
tensoring together old ones.  In fact, it has recently been found that owing
to the non-linearity of $W$ algebras, there can exist intrinsically
different types of realisations of the same $W$ algebra [6,7].  They will
lead to different string theories.  Thus it is interesting to explore all
the possible realisations for the $W$ algebras. One way to obtain
realisations is by the free-field construction {\it via} the Miura
transformation [8].  The Miura transformations constructed hitherto have
been restricted to the classical Lie algebras $A_n$, $D_n$ and $B_n$, which
give rise to realisations for $W\!A_n$, $W\!D_n$ and $W\!B_n$ algebras
respectively [9,10,11].  The $W\!A_n$ algebra is also known as $W_{n+1}$.

      First we consider $W\!A_n$ and $W\!D_n$, based on the classical
simply-laced Lie algebras $A_n$ and $D_n$.  The corresponding Miura
transformations give rise to realisations in terms of $n$ free scalars
$\vec\varphi^{(n)}\equiv (\varphi_1, \ldots, \varphi_n)$.  It was noticed in
[12,1,2,4] that one special free scalar, say $\varphi_1$, appears in the
currents of the corresponding $W$ algebra only {\it via} its energy-momentum
tensor. The contribution from $\varphi_1$ can then be replaced by an
arbitrary energy-momentum tensor with the same central charge.  This leads
to realisations of $W\!A_n$ and $W\!D_n$ in terms of an arbitrary
energy-momentum tensor, together with $(n-1)$ free scalars.  In [6] new
realisations were obtained from the Miura transformation for $n \ge 3$ in terms
of two arbitrary commuting energy-momentum tensors with the same central
charge, together with $(n-2)$ free scalars.

       In this paper we shall review how these realisations can be obtained
from Miura transformations by using certain specific reduction procedures,
{\it i.e.} by expressing the currents of $W\!A_n$ or $W\!D_n$ in terms of
those of $W\!A_{n-1}$ or $W\!D_{n-1}$ respectively, together with an extra
free scalar field.  We shall then show that these reductions are special
cases of more general reductions, which lead to realisations in terms of
multiple numbers of arbitrary commuting energy-momentum tensors with the
same central charge, together with a certain number of free scalars. It was
observed in [7] by using the free-field construction and screening charges
that such general reductions are possible in principle. An example of a
realisation of $W\!A_4$ in terms of $W\!A_2$ and $W\!A_1$ together with an
extra free scalar was given in [7].  In this paper we shall
straightforwardly derive the general reductions directly from Miura
transformations and give explicit formalae for all cases.

       Secondly, we generalise the results to $W\!B_n$
algebras, obtaining realisations in terms of a super energy-momentum
tensor and bosonic energy momentum tensors, together with some necessary
free scalars.  In all the cases $W\!A_n$, $W\!D_n$ and $W\!B_n$, the
essential characteristic of the reductions obtained in this paper is that
one can remove any desired vertex in the Dynkin diagram for $A_n$, $D_n$ or
$B_n$, and realise the associated $W$ algebra in terms of the two
commuting $W$ algebras corresponding to the two factors in the product
subalgebra.  At the end of the paper, we comment on the applications
of these realistions to $W$ strings.

\bigskip
\noindent{\it 2) Review of Previous Results}
\bigskip

      $W\!A_n$ algebras are generated by primary currents of spins
$3,4,\ldots,(n+1)$, together with an energy-momentum tensor.  A realisation
of the $W\!A_n$ algebra in terms of $n$ free scalars
$\vec\varphi^{(n)}\equiv (\varphi_1,\ldots,\varphi_n)$ is given by the Miura
transformation for $A_n\equiv su(n+1)$ [9]
$$
\prod^{n+1}_{k=1} \big (\a \partial + \vec h^{(n)}_k \cdot (\partial
\vec\varphi^{(n)})\big )=(\a \partial)^{n+1} + \sum^{n+1}_{\ell=2} W^{(n)}_\ell
(\a \partial)^{n+1-\ell}, \eqno(1)
$$
where the $\vec h^{(n)}_k$ are $n$-component vectors satisfying
$$
\eqalign{
\vec h^{(n)}_i \cdot \vec h^{(n)}_j &=\delta_{ij}- {1 \over {n+1}}\ . \cr
\sum^{n+1}_{i=1} \vec h^{(n)}_i &=0\ . \cr} \eqno(2)
$$
The quantities $W^{(n)}_\ell$ in (1) with $2 \le \ell \le (n+1)$ are
spin-$\ell$ currents that generate the $W\!A_n$ algebra.  These higher-spin
currents are not yet primary with respect to the energy-momentum tensor
$T\equiv W^{(n)}_2$, but they can be made so by adding composites and
derivatives of the lower-spin currents. The closure of the algebra only
requires that the vectors $\vec h^{(n)}_i$ satisfy the conditions (2), for
which many solutions exist.  If one just considers the
$n$-scalar realisation, all these solutions are equivalent; an orthonormal
transformation of the free scalars $\vec \varphi^{(n)}$ will map one solution
to
another.  However, we shall see later that certain choices of $\vec
h^{(n)}_i$ will make it possibe to express the currents of the $W\!A_n$
algebra in terms of the currents of arbitrary $W\!A_k$ and $W\!A_{n-k-1}$
algebras with $0 \le k \le (n-1)$, together with one extra scalar field.

      In [2,4] a specific choice for the vectors $\vec h^{(n)}_i$ is
proposed for which they have the nice property that one can
express $\vec h^{(n)}_i$ in terms of $\vec h^{(n-1)}_i$, {\it viz.}
$$
\eqalign{
\vec h^{(n)}_i &=\Big ( \vec h^{(n-1)}_i, {1 \over \sqrt{n(n+1)}}\Big ),
\qquad 1 \le i \le n \ ,\cr
\vec h^{(n)}_{n+1} &=\Big (\underbrace{0, \ldots,0}_{n-1}, -{ n  \over
\sqrt{n(n+1)} } \Big ) \ , \cr} \eqno(3)
$$
where the $(n-1)$-component vectors $h^{(n-1)}_i$, also satisfying (2) with
$n \rightarrow (n-1)$, are those for $W\!A_{n-1}$.  Substituting (3) into
(1), the left-hand side of equation (1) can be rewritten as
$$
\eqalign{
&\Big (\a\partial - n (\partial \phi_n) \Big )\prod^{n}_{\ell=1}
\Big (\a \partial +
\vec h^{(n-1)}_{\ell} \cdot (\partial \vec\varphi^{(n-1)}) + (\partial
\phi_n)\Big ) \cr
&=\Big (\a\partial - n (\partial \phi_n)\Big ) e^{-\phi_n/\a}
\prod^{n}_{\ell=1}\Big (\a \partial +
\vec h^{(n-1)}_{\ell} \cdot (\partial \vec\varphi^{(n-1)})\Big )
e^{\phi_n/\a} \ , \cr} \eqno(4)
$$
where we have defined $\phi_n \equiv {1 \over \sqrt{n(n+1)}} \varphi_n$.
Using the Miura transformation (1) for $W\!A_{n-1}$, we can write the
$\prod_{\ell=1}^{n}$ factors in the right-hand side of the equation (4) in
terms of the currents of the $W\!A_{n-1}$ algebra.  Since the currents of
$W\!A_{n-1}$, differential polynomials of free scalars
$\vec\varphi^{(n-1)}$, commute with $\varphi_n$, we can realise the currents
of $W\!A_n$ in terms of those of an arbitrary $W\!A_{n-1}$ and an extra
scalar $\varphi_n$.  Applying this procedure recursively leads to a
realisation of the $W\!A_n$ algebra in terms of an arbitrary energy-momentum
tensor and $(n-1)$ additional scalar fields $(\varphi_2, \ldots,
\varphi_n)$.

     In [5] a similar reduction procedure is proved for the $W\!D_n$ algebras.
One can realise the currents of the $W\!D_n$ algebra in terms of those of
$W\!D_{n-1}$ and an extra free scalar.  Since $D_2 \cong A_1\times A_1$, the
algebra of $W\!D_2$ is isomorphic to the direct product of two independent
Virasoro algebras but with the same central charge.  Exploiting the fact that
$A_3 \cong D_3$ and hence $W\!A_3 \cong W\!D_3$, one can therefore realise
$W\!A_n$ and $W\!D_n$ for $n \ge 3$ in terms of either one arbitrary
energy-momentum tensor together with $(n-1)$ additional free scalars [5] or two
arbitrary energy-momentum tensors together with $(n-2)$ additional free
scalars [6].

\bigskip
\noindent{\it 3) General Reduction for $W\!A_n$ }
\bigskip

       The reduction $W\!A_n\to W\!A_{n-1}$ described above is dictated by the
specific choice of the vectors $\vec h^{(n)}_i$ given in (3).  This choice
can be easily generalised so that the vectors $\vec h^{(n)}_i$ can be
expressed in terms of $\vec h^{(n-k-1)}_i$ and $\vec h^{(k)}_i$, with $0 \le
k \le (n-1)$, {\it viz.}
$$
\eqalign{
\vec h^{(n)}_i&=\Big ( \vec h^{(n-k-1)}_i,\,\,\underbrace{0,\ldots,0}_k,\,\,
\sqrt{\ft{k+1}{(n-k)(n+1)}}\, \Big )\ , \qquad 1 \le i \le (n-k) \ , \cr
\vec h^{(n)}_j&=\Big ( \underbrace{0,\ldots,0}_{n-k-1},\,\,
\vec h^{(k)}_{j+k-n}, \,\,-\sqrt{\ft{n-k}{(k+1)(n+1)}}\, \Big )\ ,
\qquad (n-k+1) \le j \le (n+1)\ ,\cr} \eqno(5)
$$
where $\vec h^{(n-k-1)}_i$ and $\vec h^{(k)}_i$  are the corresponding vectors
for $W\!A_{n-k-1}$ and $W\!A_k$, satisfying (2) with $n\to (n-k-1)$ and
$n\to k$  respectively.  It is easy to check that the
vectors $\vec h^{(n)}_i$ given in (5) satisfy the condition (2).  In
particular, when $k=0$, equation (5) will reduce to the special case given
in equation (3). Substituting (5) into the Miura transformation (1), the
left-hand side of the equation reads
$$
\eqalign{
&\prod^{n+1}_{\ell=n-k+1} \Big ( \a \partial + \vec h^{(k)}_{\ell+k-n} \cdot
(\partial \widetilde\varphi^{(k)}) -(n-k) (\partial
\phi_n) \Big )\cr
&\times\prod^{n-k}_{\ell=1} \Big ( \a \partial + \vec
h^{(n-k-1)}_{\ell} \cdot (\partial \vec \varphi^{(n-k-1)}) +
(k+1)(\partial \phi_n) \Big ) \cr
&=e^{(n-k)\phi_n/\a} \prod^{n+1}_{\ell=n-k+1} \Big ( \a \partial +
\vec h^{(k)}_{\ell+k-n} \cdot (\partial \widetilde\varphi^{(k)}) \Big )
e^{-(n-k)\phi_n/\a} \cr
&\times e^{-(k+1)\phi_n/\a} \prod^{n-k}_{\ell=1} \Big ( \a \partial + \vec
h^{(n-k-1)}_{\ell} \cdot (\partial \vec \varphi^{(n-k-1)})\Big )
e^{(k+1)\phi_n/\a}\ , \cr} \eqno(6)
$$
where we have defined $\phi_n\equiv\ft{1}{\sqrt{(k+1)(n-k)(n+1)}}\,\varphi_n$
and we also have split $\vec\varphi^{(n)}$ into
$$
\vec \varphi^{(n)}=\big (\vec \varphi^{(n-k-1)}, \,\, \widetilde
\varphi^{(k)}, \,\, \varphi_n \big ) \eqno(7)
$$
with $\widetilde\varphi^{(k)} \equiv (\varphi_{n-k},\ldots, \varphi_{n-1})$.
Using the Miura transformation (1), we can write the $\prod_{\ell=1}^{k}$
factors in the right-hand side of equation (6) in terms the currents of
$W\!A_k$ and write the $\prod_{\ell=n-k+1}^{n+1}$ factors in terms of the
currents of $W\!A_{n-k-1}$.  We therefore obtain a realisation for the
$W\!A_n$ algebra in terms of commuting $W\!A_k$ and $W\!A_{n-k-1}$
realisations with an additional scalar $\varphi_n$.  The integer $k$ takes
values $0,\ldots, (n-1)$. When $k=0$ or $k=n-1$, we will recover the result
given in (4).  This reduction can be understood schematically from the
Dynkin diagram
$$
\overbrace{\circ\!\!-\!\!\!-\!\!\!-\!\!\circ\!\!-\!\!\! -\!\!\!
-\cdots-\!\!\!-\!\!\!-\!\!\circ}^{A_k\to W\!A_k}\!\!-\!\!\!-\!\!\!
\overbrace{-\!\!\otimes\!\!-}^{\partial\varphi}\!\!\!-
\!\!\!-\!\!\overbrace{\circ
\!\!-\!\!\!-\!\!\!-\!\!\circ\!\!-\!\!\!-\!\!\!-\cdots-\!\!\!-\!\!\!
-\!\!\circ}^{A_{n-k-1}\to W\!A_{n-k-1}} \ .\eqno(8)
$$
One can obtain a subalgebra by removing any vertex in the Dynkin diagram for
$A_n$; consequently, one can realise the $W\!A_n$ in terms of the
corresponding lower $W$ algebras based on the direct-product subalgebra with
an extra scalar corresponding to the vertex removed.  Applying this
reduction procedure recursively, one can realise $W\!A_n$ algebra in terms
of $k$ energy-momentum tensors with $0 \le k \le [\ft{n+1}{2}]$, together
with $(n-k)$ additional scalars.  Since the energy-momentum tensors commute
with each other and with the additional scalars, they can be arbitrary and
independent of one another.  However the central charges for all these
energy-momentum tensors must be same.  This follows from the fact that the
vectors $\vec h^{(1)}_i$ for $W\!A_1\equiv {\rm Virasoro}$ are unique.
Consequently, the background charges for these energy-momentum tensors,
originally derived from the Miura transformation, all take the same value
$\a/\sqrt{2}$, which leads to the same central charge $c^{\rm eff}$ given by
$$
c^{\rm eff}=1+6 \a^2 \ .\eqno(9)
$$
The total central charge of the $W\!A_n$ algebra is independent of which of the
above realisations is used.  It is given by
$$
\eqalign{
c(W\!A_n)&=n+12 \a^2 (\vec \rho)^2 \cr
&\equiv n+ 12 \a^2 \Big (\sum^{n+1}_{j=2} (1-j) \vec h^{(n)}_j \Big ) ^2
\cr
&=n\Big ( 1+(n+1)(n+2)\a^2 \Big )\ .\cr}\eqno(10)
$$

\bigskip
\noindent{\it 4) General Reduction for $W\!D_n$ }
\bigskip

       Now we turn our attention to the $W\!D_n$ algebras, which are
generated by currents $W^{(n)}_{2j}(z)$ of spin $s=2j$, with $j=1,\ldots,
(n-1)$, and a current $U^{(n)}(z)$ of spin $n$.  A similar reduction also
works in this case.  One can realise the $W\!D_n$ in terms of $W\!A_k$
and $W\!D_{n-k-1}$ and an extra scalar, with $0 \le k \le (n-1)$.  To see
this, we write down the $n$-scalar realisation from the Miura transformation
for the $W\!D_n$ algebra. The spin-$n$ current is given by the Miura-type
transformation [10,11]
$$
U^{(n)}(z)=\prod^{n}_{\ell=1} \big (\a \partial - \vec \sigma^{(n)}_\ell
\cdot (\partial \varphi^{(n)}) \big )\cdot\oneone\ ,\eqno(11)
$$
where the $\vec \sigma^{(n)}_\ell$ are $n$-component vectors satisfying
$$
\vec \sigma^{(n)}_i \cdot \vec \sigma^{(n)}_j =\delta_{ij}\ . \eqno(12)
$$
The $W^{(n)}_{2j}$ currents can then be read off from the operator-product
expansion $U^{(n)}(z)U^{(n)}(w)$:
$$
U^{(n)}(z)U^{(n)}(w)\sim {a_n \over (z-w)^{2n}} + \sum^{n-1}_{j=1} {a_{n-j}
\over (z-w)^{2n-2j}} \big ( W^{(n)}_{2j}(z) +W^{(n)}_{2j}(w)\big )\ ,
\eqno(13)
$$
where $a_j$ are normalisation constants given by $a_j=(-1)^{j+1}
\prod_{\ell=1}^{j-1} \big (1+2\ell(2\ell+1)\a^2 \big)$.

      The closure of the $W\!D_n$ algebra requires that the vectors $\vec
\sigma^{(n)}_i$ satisfy the condition (12).  In [10,11] the simplest solution
to (12) was used, namely $\vec \sigma^{(n)}_i=(0,\ldots,0,1,0,\ldots,0)$,
where the $1$ occurs in the $i$-th entry. This solution enables us to
express the currents of the $W\!D_n$ algebra in terms of those of
$W\!D_{n-1}$ and an extra scalar [5].  In general, one can achieve
condition (12) by expressing the $\vec \sigma^{(n)}_i$ in terms of $\vec
\sigma^{(n-k-1)}_i$ and $\vec h^{(k)}_i$, {\it viz.}
$$
\eqalign{
\vec \sigma^{(n)}_i&=\Big (\vec \sigma^{(n-k-1)}_i,\,\, \underbrace{0,
\ldots, 0}_{k},\,\, 0 \Big)\ ,\qquad 1 \le i \le (n-k-1) \ ,\cr
\vec \sigma^{(n)}_j&=\Big (\underbrace{0,\ldots,0}_{n-k-1}, \,\, \vec
h^{(k)}_{j-n+k+1}, \,\, \ft1{\sqrt{k+1}} \Big )\ ,\qquad (n-k) \le j \le n \ ,
\cr } \eqno(14)
$$
where $\vec\sigma^{(n-k-1)}_i$ and $\vec h^{(k)}_j$ are the vectors
for $W\!D_{n-k-1}$ and $W\!A_k$, satisfying (2) and (12) respectively.
Substituting (14) into (11), we can write the spin-$n$ current $U^{(n)}$ in
terms of the spin-$(n-k-1)$ current $U^{(n-k-1)}$ of $W\!D_{n-k-1}$ and the
currents of $W\!A_k$, together with an extra scalar.  Consequently, it
follows from equation (14) that we can express all the currents of $W\!D_n$
in terms of those of $W\!D_{n-k-1}$ and $W\!A_k$ together with the extra
field.  This reduction procedure can be summarised by the Dynkin diagram
$$
\overbrace{\circ\!\!-\!\!\!-\!\!\!-\!\!\circ\!\!-\!\!\! -\!\!\!
-\cdots-\!\!\!-\!\!\!-\!\!\circ}^{A_k\to W\!A_k}\!\!-\!\!\!-\!\!\!
\overbrace{-\!\!\otimes\!\!-}^{\partial\varphi}\!\!\!-
\!\!\!-\!\!\overbrace{\circ
\!\!-\!\!\!-\!\!\!-\!\!\circ\!\!-\!\!\!-\!\!\!-\cdots-\!\!\!-\!\!\!
-\!\!\circ\!\!-\!\!\!-\!\!\!-\!\!\circ\!{\Bigg\langle\!}^{\displaystyle{\circ}}
_{\displaystyle{\circ}}
}^{D_{n-k-1}\to W\!D_{n-k-1}}\ .\eqno(15)
$$
Applying this reduction, and the reduction for $W\!A_n$, recursively gives
rise to realisations of the $W\!D_n$ algebra in terms of $k$ energy-momentum
tensors together with $(n-k)$ free scalars, with $0 \le k \le
[\ft{n+1}{2}]$.  Like the case of $W\!A_n$, these energy-momentum tensors
are arbitrary and independent, but with the same central charge given in
(9).  The total central charge is given by
$$
c(W\!D_n)=n \Big(1+(2n-1)(2n-2)\a^2 \Big ).\eqno(16)
$$
Note that taking $k=n-1$ in the above reduction defined by (14) corresponds
to reducing $W\!D_n$ to $W\!A_{n-1}$, represented by the deletion of one of
the ``ears'' in the Dynkin diagram (15).  When $k=n-2$, the corresponding
subalgebra of $D_n$ is $A_{n-2}\times D_1$.  Although $D_1$ is isomorphic to
$A_1$, the $W\!D_1$ algebra, as defined by the Miura transformation (11), is
not isomorphic to $W\!A_1\cong {\rm Virasoro}$; instead it is an algebra
with just a spin-1 current $\del\varphi_1$.  Thus when $k=n-2$ we get
$W\!A_n$ realised in terms of $W\!A_{n-2}$ and 2 additional free scalars.

\bigskip
\noindent{\it 5) General Reduction for $W\!B_n$ }
\bigskip

       Having obtained new realisations for bosonic $W\!A_n$ and $W\!D_n$
based on the classical simply-laced Lie algebras $A_n$ and $D_n$, we should
like to study the case of $W\!B_n$, which is based on the classical
non-simply-laced Lie algebra $B_n$.  The $W\!B_n$ algebra has bosonic
currents of spins $s=2,4,\ldots,2n$, and an additional fermionic current
with spin $(n+\ft12)$.  The special case of $W\!B_1$ is linear, and in fact
is simply the $N=1$ super-Virasoro algebra.  The higher-$n$ $W\!B_n$
algebras can thus be thought of as higher-spin extensions of the $N=1$
super-Virasoro algebra; as such, they may turn out to be the most suitable
candidates for generating higher-spin extensions of superstring theory.
The first non-trivial ({\it i.e.}\ non-linear) example, $W\!B_2$, was
constructed explicitly in [13]. The fermionic spin-$(n+\ft12)$ current
$Q^{(n)}(z)$ of $W\!B_n$ plays an analogous r\^ole to the bosonic spin-$n$
current $U^{(n)}(z)$ in the $W\!D_n$ algebra.  It is given by the Miura-type
transformation [10]
$$
Q^{(n)}(z)=\Big (\prod^{n}_{\ell=1} \big (\a \partial - \vec
\sigma^{(n)}_\ell
\cdot (\partial \varphi^{(n)}) \big ) \Big )\psi\ ,\eqno(17)
$$
where the $\vec \sigma^{(n)}_i$ are $n$-component vectors satisfying
condition (12) and $\psi$ is a real free fermion field.  The bosonic
currents $W^{(n)}_{2\ell}$ can be read off from the operator-product expansion
$Q^{(n)}(z) Q^{(n)}(w)$:
$$
Q^{(n)}(z) Q^{(n)}(w)\sim {b_n \over (z-w)^{2n+1}} + \sum^n_{j=1} {b_{n-j}
\over (z-w)^{2n+1-2j} }\big ( W^{(n)}_{2j}(z) + W^{(n)}_{2j}(w) \big )
\eqno(18)
$$
with $b_j=\prod_{\ell=1}^j \big(1+2\ell(2\ell-1)\a^2\big)$.  Like the case of
$W\!D_n$, we can express the vectors $\vec \sigma^{(n)}_i$ in terms of $\vec
\sigma^{(n-k-1)}_i$ and $\vec h^{(k)}_j$ given in (14).  Thus we can realise
the $W\!B_n$ algebra in terms of $W\!B_{n-k-1}$ and $W\!A_k$ together with
an extra scalar, with $ 0 \le k \le (n-1)$.  The result for the case $k=0$
was first obtained in [5].  Schematically, the general case is summarised by
the Dynkin diagram
$$
\overbrace{\circ\!\!-\!\!\!-\!\!\!-\!\!\circ\!\!-\!\!\! -\!\!\!
-\cdots-\!\!\!-\!\!\!-\!\!\circ}^{A_k\to W\!A_k}\!\!-\!\!\!-\!\!\!
\overbrace{-\!\!\otimes\!\!-}^{\partial\varphi}\!\!\!-
\!\!\!-\!\!\overbrace{\circ
\!\!-\!\!\!-\!\!\!-\!\!\circ\!\!-\!\!\!-\!\!\!-\cdots-\!\!\!-\!\!\!
-\!\!\circ\!\!\!=\!\!\!=\!\!\!=\!\!\!\bullet}^{B_{n-k-1}\to W\!B_{n-k-1}}\
.\eqno(19)
$$
Applying this reduction procedure and the reduction for $W\!A_n$
recursively, one can realise the $W\!B_n$ algebra in terms of $W\!B_1\cong$
super-Virasoro, $k$ arbitrary energy-momentum tensors, and $(n-k-1)$
additional scalars, with $0 \le k \le [\ft{n-1}{2}]$.  The total central
charge for $W\!B_n$ is given by
$$
c(W\!B_n)=(n+\ft12)(1+2n(2n-1)\a^2)\ . \eqno(20)
$$
The central charges for each arbitrary energy-momentum tensor are again the
same and given by (9).  The central charge $c^{\rm eff}_{\rm sup}$ for the
super energy-momentum tensor is
$$
c^{\rm eff}_{\rm sup}=\ft32+3\a^2 \ . \eqno(21)
$$
Taking $k=n-1$ in (14) corresponds here to the reduction of $W\!B_n$ to
$W\!A_{n-1}$, represented by deleting the black vertex in the Dynkin diagram
(19).  In this case there is no super energy-momentum tensor, and $W\!B_n$
is realised in terms of the currents of $W\!A_{n-1}$ together with a free
scalar and a free fermion.

\bigskip
\noindent{\it 6) Application in String Theories }
\bigskip

      String theory is two-dimensional gravity coupled to a critical matter
system that includes free scalar fields which are interpreted as coordinates
on the target spacetime.  One may construct generalisations of
two-dimensional gravity, by gauging matter systems with $W$ symmetries.  If
the matter systems are critical, and include free scalars, then one obtains
$W$-string theories.  The realisations for $W$ algebras obtained above all
have two types of elements: some arbitrary energy-momentum tensors (or in
the case of $W\!B_n$, energy-momentum tensors and a super energy-momentum
tensor) and some necessary additional free scalars. Calculations for
specific examples [1-6]
show that the physical-state conditions for higher-spin currents just serve the
purpose of ``freezing'' these additional coordinates, {\it i.e.} the momenta
conjugate to these coordinates are frozen to certain specific values by the
physical state conditions.  If any of these (super) energy-momentum tensors
comprises $D$ free scalars, these scalars form the $D$-dimensional
physically-observable coordinates on the target spacetime. Thus a $W$-string
theory is effectively described by a set of independent ordinary (super)
Virasoro-like strings with a non-standard effective central charge and a set
of non-standard spin-$2$ intercept values.  We shall solve these central
charges and exhibit a numerological connection with (super) Virasoro minimal
models, for the $W\!A_n$, $W\!D_n$ and $W\!B_n$ strings.

      Anomaly freedom of a $W$ string requires that the total central charge
take a certain specific value in order to cancel the anomalous
contributions from the ghosts.  The critical central charges for $W$ strings
are given by [5]
$$
\eqalign{
c^*&=2n(2n^2+6n+5)\ , \qquad\qquad {\rm for}\ \  W\!A_n \ ,\cr
c^*&=2n(8n^2-12n+5) \ ,\qquad\qquad {\rm for}\ \  W\!D_n \ ,\cr
c^*&=(2n+1)(8n^2-4n+1) \ ,\qquad {\rm for}\ \  W\!B_n \ .\cr
}\eqno(22)
$$
Substituting these values into (10), (16) and (20), one can solve for the
critical background-charge parameter $\a^*$ for $W\!A_n$, $W\!D_n$ and
$W\!B_n$ respectively.  One can then, from (9) and
(21), obtain the effective central charges for the corresponding $W$
strings:
$$
\eqalign{
&\left.
\eqalign{
c^{\rm eff}&=26-\Big (1-{6 \over (n+1)(n+2)} \Big )\ ,\qquad {\rm for}\ \
W\!A_n \ ,\cr
c^{\rm eff}&=26-\Big (1-{6\over (2n-2)(2n-1)} \Big )\ ,\qquad {\rm for}
\ \ W\!D_n \ ,\cr}\right.\cr
&\left.
\eqalign{
c^{\rm eff}&=26-\Big (1-{6\over 2n(2n-1)} \Big)\ ,\cr
c^{\rm eff}_{\rm sup}&=15-\Big (3/2-{12 \over 4n(4n-2)} \Big )\ ,\cr}
\right\}\qquad {\rm for }\ \  W\!B_n\ .\cr}\eqno(23)
$$
These effective central charges are all equal to the critical central charge
of the (super) Virasoro string minus the central charges of certain unitary
(super) Virasoro minimal models.  Such connections between $W$ strings and
corresponding unitary minimal models are strengthened by the fact that if
one writes the effective spin-$2$ intercepts $L^{\rm eff}_0=1- L^{\rm
min}_0$ for $W\!A_n$ and $W\!D_n$, then the values of $L^{\rm min}_0$ are
precisely the dimensions of primary fields of the corresponding minimal
models.  In the case of $W\!B_n$, one similarly writes and $L^{\rm
eff}_0=1-L^{\rm min}_0$ for the energy-momentum tensors, and $(L^{\rm
eff}_{\rm sup})_0=\ft12-L^{\rm min}_0$ for the super energy-momentum tensor.
It has been shown in [4,5,6] that connections with minimal models occur in
the $W$ strings with realisation in terms of one (or two) arbitrary
energy-momentum tensors for $W\!A_n$ and $W\!D_n$, and one super
energy-momentum tensor for $W\!B_n$.  We may expect that this connection
exists also in the more general realisations obtained here.

\bigskip
\noindent{\it 7) Summary}
\bigskip

     In this paper, we have studied general classes of realisations for $W$
algebras.  The known realisations all originate from Miura transformations.
The Miura transformations for the $W\!A_n$, $W\!D_n$ and $W\!B_n$ algebras
all share the common feature that they can be ``factorised'' as a product of
two Miura transformations for two smaller commuting $W$ algebras.
Specifically, the possible factorisations are dictated precisely by the
canonical subalgebras of the underlying $A_n$, $D_n$ or $B_n$ Lie algebras
obtained by removing any vertex from the corresponding Dynkin diagram.  Thus
for example the currents of $W\!A_n$ can be realised in terms of those for
$W\!A_p$ and $W\!A_q$ algebras such that $p+q=n-1$, together with an
additional free scalar field (corresponding to the deleted vertex).
Similarly, $W\!D_n$ may be realised in terms of $W\!D_p$ and $W\!A_q$, and
$W\!B_n$ may be realised in terms of $W\!B_p$ and $W\!A_q$, with $p+q=n-1$ in
each case.  Again, there is one additional free scalar in each realisation.

     A straightforward extension of the above results is to apply the procedure
recursively, so that one is effectively deleting more non-adjacent vertices
in the original Dynkin diagram.  The maximal case is when one deletes all
alternate vertices.  One can therefore obtain realisations of $W\!A_n$ or
$W\!D_n$ in terms of $\ell$ mutually-commuting energy-momentum tensors and
$(n-\ell)$ additional free scalars, for any $\ell$ in the range $0\le \ell\le
[\ft{n+1}{2}]$.  In the non-simply-laced case of $W\!B_n$, one can realise
it in terms of an $N=1$ super energy-momentum tensor, $\ell$ bosonic
energy-momentum tensors, and $(n-\ell-1)$ additional scalars, for any $\ell$ in
the range $0\le \ell\le [\ft{n-1}{2}]$.  Alternatively, in view of the
remark at the end of section 5, it can be realised in terms of $\ell$
energy-momentum tensors, $(n-\ell-1)$ free scalars and a free fermion, for
$\ell$ in the range $0\le\ell\le[\ft{n}2]$.

      An application of these realisations is to the corresponding
$W$-string theories.  Results for the previously-known realisations of
$W\!A_n$ or $W\!D_n$ with $\ell=1$ and $\ell=2$, and $W\!B_n$ with $\ell=0$,
have revealed connections with certain unitary minimal models.  It appears,
by looking at the central charges for the effective energy-momentum tensors
appearing in the more general realisations considered here, that these
connections should persist for all the new realisations.  There is an
accumulation of evidence that all $W$-string theories are related to
Virasoro-type strings and minimal models.  The underlying significance of
this relation remains to be understood.

\np

\centerline{\bf REFERENCES}
\frenchspacing
\bigskip

\item{[1]}C.N.\ Pope, L.J.\ Romans and K.S.\ Stelle, {\sl Phys.\
Lett.}\ {\bf 268B} (1991) 167;\nl
{\sl Phys.\ Lett.}\ {\bf 269B} (1991) 287.

\item{[2]}S.R.\ Das, A.\ Dhar and S.K.\ Rama, {\sl Mod.\ Phys.\ Lett.}\
{\bf A6} (1991) 3055;\nl
``Physical states and scaling properties of $W$  gravities and $W$ strings,''
TIFR/TH/91-20.

\item{[3]}C.N.\ Pope, L.J.\ Romans, E.\ Sezgin and K.S.\ Stelle,
{\sl Phys.\ Lett.}\ {\bf 274B} (1992) 298.

\item{[4]}H.\ Lu, C.N.\ Pope, S.\ Schrans and K.W.\ Xu, ``The Complete
Spectrum of the $W_N$ String,''  preprint CTP TAMU-5/92, KUL-TF-92/1.

\item{[5]}H.\ Lu, C.N.\ Pope, S.\ Schrans and X.J.\ Wang, ``On Sibling and
Exceptional $W$ Strings,''  preprint CTP TAMU-10/92, KUL-TF-92/8, to appear
in {\sl Nucl.\ Phys.\ }{\bf B}.

\item{[6]}H.\ Lu, C.N.\ Pope, S.\ Schrans and X.J.\ Wang, ``New Realisations
of $W$ Algebras and $W$ Strings,''  preprint CTP TAMU-15/92, KUL-TF-92/11.

\item{[7]}G.M.T.\ Watts, ``A Note on $W$-algebra Realisations,''  preprint
DUR-CPT 92-15.

\item{[8]}V.G.\ Drinfeld and V.V.\ Sokolov, {\sl Journ.\ Sov.\ Math.}
\ {\bf 30} (1985) 1975.

\item{[9]}V.A.\ Fateev and S.\ Lukyanov,  {\sl Int.\ J.\ Mod.\  Phys.}
\ {\bf A3} (1988) 507.

\item{[10]}S.L.\ Lukyanov and V.A.\ Fateev, {\sl Sov.\ Scient.\ Rev.}\ {\bf
A15} (1990).

\item{[11]}S.L.\ Lukyanov and V.A.\ Fateev, {\sl Sov.\ J.\ Nucl.\ Phys.}
\ {\bf 49} (1989) 925.

\item{[12]}L.J.\ Romans, {\sl Nucl. Phys.}\ {\bf B352} (1991) 829.

\item{[13]}J.M.\ Figueroa-O'Farrill, S. Schrans and K. Thielemans, {\sl
Phys.\ Lett.}\ {\bf 263} (1991) 378.

\bye